\documentclass[prb,twocolumn,showpacs]{revtex4}     
\usepackage{epsfig,amsmath} 

\begin{document}

\title{Absence of gapped broken inversion symmetry phase of electrons in bilayer graphene under renormalized ring-diagram approximation} 

\author{Xin-Zhong Yan$^{1}$ and C. S. Ting$^2$}
\affiliation{$^{1}$Institute of Physics, Chinese Academy of Sciences, P.O. Box 603, 
Beijing 100190, China\\
$^{2}$Department of Physics, University of Houston, Houston, Texas 77204, USA}
 
\date{\today}
 
\begin{abstract}
On a lattice model, we study the possible existence of a gapped broken inversion symmetry phase (GBISP) of electrons with long-range Coulomb interactions in bilayer graphene using both the self-consistent Hartree-Fock approximation (SCHFA) and the renormalized ring-diagram approximation (RRDA). The RRDA takes into account the charge density fluctuations beyond the SCHFA. Although the GBISP at low temperature and low carrier concentration is predicted by the SCHFA, we show here that this state can be substantially suppressed by the charge density fluctuations in the RRDA. We also present a numerical algorithm for calculating the self-energy of electrons with the singular long-range Coulomb interaction on the lattice model. 
\end{abstract}


\maketitle
\section{introduction}

Because of its tunable band gap, which can be changed through an external gate voltage, bilayer graphene is a promising material with a great potential for application to new electronic devices.\cite{Ohta,Oostinga,McCann,Castro} In the low-carrier-doping regime of bilayer graphene, electrons are strongly coupled via Coulomb interactions. The phase of bilayer graphene in this low carrier concentration, and at low temperature, is still not completely understood. Several candidates have been suggested for the ground state, such as a ferroelectric-layer asymmetric state,\cite{Min,Nandkishore,Zhang,Jung,MacDonald} a layer-polarized antiferromagnetic state,\cite{Kharitonov,Vafek0} a quantum anomalous Hall state,\cite{Jung,Nandkishore1,Zhang1} a quantum spin Hall state,\cite{Jung,Zhang1} a quantum valley Hall state,\cite{Zhang2} a charge density wave state,\cite{Dahal} and the possibility of gapless states, such as the nematic state.\cite{Vafek,Lemonik} The experimental observations on the ground state of bilayer graphene, all performed on high quality suspended samples, are also controversial. Some experimental results showed that the system is gapped at the neutrality point,\cite{Weitz,Freitag,Velasco,Bao} whereas one experiment found a gapless state.\cite{Mayorov} So far, most of the theoretical studies are based on the self-consistent Hartree-Fock approximation (SCHFA),\cite{Min,MacDonald,Zhang2} many-body perturbation theory,\cite{Zhang} and the renormalization group approach.\cite{Lemonik,Vafek,Vafek0} All the above approaches have been applied to the simplified two-\cite{Min,MacDonald,Zhang2,Zhang,Lemonik,Vafek0} and four-band\cite{Vafek} continuum models. It is well known that the SCHFA usually overestimates the order parameter characterizing a broken symmetry phase and the transition temperature because it neglects the fluctuations of the effective one-body interaction field and of other one-body observables such as the charge density. Since the understanding of the electronic state of bilayer graphene at low carrier doping and low temperature is a fundamental issue for graphene physics, it is necessary to investigate the state with a more sophisticated approach that takes into account the effect of charge density fluctuations on top of the mean-field ground state.

In this work, we study the existence of a gapped broken inversion symmetry phase (GBISP) using both the SCHFA and the renormalized ring-diagram approximation (RRDA).\cite{YanB11} The RRDA takes into account the charge density fluctuation (CDF) effect beyond the mean field and satisfies the microscopic conservation laws.\cite{Baym} For an electron system with long-range Coulomb interactions, CDF is the predominant contribution to the self-energy of electrons. It has been shown\cite{YanE} that the RRDA results for the ground-state energy of two- and three-dimensional interacting electron gases are more accurate than the random-phase approximation (RPA) results when compared with Monte Carlo simulations. 

In the RRDA, the Green's function and self-energy are self-consistently determined by coupled integral equations. The self-consistent calculation of the self-energy in momentum space involves carrying out many convolutions, which are numerically expensive. In order to reduce the computational time required by our approach, we convert the convolutions in momentum space to multiplications in real space. Since the continuum model is the low-energy limiting case of the lattice model, the momentum of the electrons is confined within two valleys around the Dirac points.\cite{McCann2,Novoselov} Because of the finite momentum cutoff for each valley, the conversion of the convolution from momentum space to real space is no longer valid for the two- and four-band continuum models. Instead of modeling bilayer graphene with an effective continuum model, we therefore sketch it as a bilayer of a hexagonal lattice model. The lattice model does not require a momentum-space cutoff, and is therefore immune to the aforementioned problems of the continuum models.

The key problem in calculating the self-energy is to manage to deal with the long-range Coulomb interaction between electrons accurately. For the two-dimensional system under consideration, this interaction is inversely proportional to the momentum transfer $q$ in the long-wavelength limit. In a continuum model, one can transform the $1/q$ singularity to the logarithmic one after performing the azimuthal integration\cite{YanB72} and then get rid of the logarithmic singularity by special treatment. In a lattice model, however, we cannot perform the azimuthal integration analytically and must face the $1/q$ singularity. Since dealing with the long-range Coulomb interaction is inevitable in many-body problems, we now present a numerical algorithm to tackle the interaction divergence issues systematically. 

\section{Lattice model}

The lattice structure of bilayer graphene is shown in Fig. 1. The unit cell in each layer is represented by a diamond. The unit cell of the bilayer system contains four atoms denoted as a$_1$, b$_1$, a$_2$ and b$_2$. The lattice constant of monolayer graphene is defined as the distance between two nearest corner atoms in the diamond and is given by $a \approx 2.4$ \AA~. The interlayer distance is $z_0 = 3.34$ \AA~ $\approx 1.4 a$. The energy of electron hopping between the nearest-neighbor (NN) carbon atoms in each layer is $t \approx 2.82$ eV,\cite{Bostwick} while the interlayer NN hopping is $t_1 \approx 0.39$ eV.\cite{Misu} 

The Hamiltonian describing the electrons is given by
\begin{eqnarray}
H=-\sum_{ij\sigma}t_{ij}c^{\dagger}_{i\sigma}c_{j\sigma}+\frac{1}{2}\sum_{ij}\delta n_iv_{ij}\delta n_j \label {hm}
\end{eqnarray}
where $c^{\dagger}_{i\sigma}$ creates an electron at site $i$ with spin $\sigma$, $\delta n_j=n_j-n$ with $n_j$ as the electron density operator at site $j$ and $n$ the average occupation number of electrons per site (which is also the charge number of the neutralizing background), and $v_{ij}$ is the Coulomb interaction between electrons at sites $i$ and $j$. The model is restricted to NN hopping within the same layer and between the adjacent sites on top and bottom layers as shown in Fig. 1. As described by Eq. (\ref{hm}), we consider here only the charge-charge interactions. Since the long-range antiferromagnetic order is prohibited\cite{Mermin} in two-dimensional space, we neglect the antiferromagnetic coupling due to the on-site repulsion in the present work.

We now consider the behavior of Coulomb interaction $v_{ij}$ between two electrons at sites $i$ and $j$. At long distance, $v_{ij}$ is given by $v_{ij} = e^2/\epsilon r_{ij}$ with $\epsilon$ the dielectric constant in the high-frequency limit of the system and $r_{ij}$ the distance. However, at short distances, because of the spread of the $\pi$-orbital wave function of the conduction electrons, $v_{ij}$ is weakened from the behavior $1/r_{ij}$. Taking the effect of the wave function spread into account, we model the interaction as
\begin{eqnarray}
v_{ij}=\frac{e^2}{\epsilon r_{ij}}[1-\exp(-r_{ij}/r_0)] \label{int}
\end{eqnarray}
with $r_0 = a$. Clearly, $v_{ij}$ behaves as $e^2/\epsilon r_{ij}$ at large $r_{ij}$, while it is suppressed from the `bare' Coulomb interaction ($e^2/\epsilon r_{ij}$) at small $r_{ij}$. In particular, at $r_{ij} = 0$, it is given by a finite value $e^2/\epsilon r_0$. For the present electron system with long-range Coulomb interactions, the final result under consideration should not be sensitively dependent upon the details of the short-range behavior of the interaction. This can be understood from the behavior of its Fourier component in momentum space. The Fourier component is singular at the long-wavelength limit and the singular part is independent of the short-range behavior. At low carrier concentration, the electrons state is mainly determined by the singular part of the interaction. We use the dimensionless constant $g \equiv e^2/\epsilon at$ to denote the strength of Coulomb coupling. The range $0.4 \le g < 1.8$ covers the cases of various experimental setups, from suspended bilayer graphene (BLG) to BLG placed on substrates\cite{Jang} as SiO$_2$ and ice.
 
\begin{figure} 
\centerline{\epsfig{file=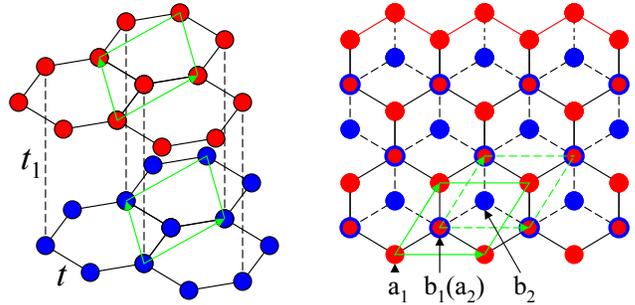,width=9. cm}}
\caption{Left: Structure of Bernal stacking bilayer graphene. Right: Top view of the bilayer graphene. The parameters $t$ and $t_1$ are the electron hopping energies between one atom and its nearest neighbor belonging to the same layer, and to the neighboring layer above or below, respectively. The unit cell of each layer is represented by the green-sided diamond. } 
\end{figure} 

The system defined by Eq. (\ref{hm}) satisfies the particle-hole symmetry. To see this, we denote the doped electron concentration per carbon atom as $\delta$ and have $n = 1+\delta$. Under the transformation $\delta \to -\delta$ and $c_{j,\sigma} \rightarrow +(-) c^{\dagger}_{j,\sigma}$ and $c^{\dagger}_{j,\sigma} \rightarrow +(-) c_{j,\sigma}$ for electrons at a$_j$ (b$_j$) sites, $H$ is unchanged. Furthermore, $K = H -\mu(\hat N - N_0)$ ($\hat N$ being the total electron number operator  and $N_0$ being the total number of lattice sites, so that the operator $N - N_0$ refers to the total number of doped electrons) is also unchanged under the above electron-hole transformation, provided $\mu \to -\mu$. Thus, the chemical potential $\mu$ must be an odd function of $\delta$.

The Green's function $G$ of the electron system is defined as 
\begin{eqnarray}
G(i,j,\tau-\tau') = -\langle T_{\tau}C_{i\sigma}(\tau)C^{\dagger}_{j\sigma}(\tau')\rangle \label {gf}
\end{eqnarray}
where $C^{\dagger}_{j\sigma} = (c^{\dagger}_{a_1j\sigma},c^{\dagger}_{b_1j\sigma},c^{\dagger}_{a_2j\sigma },c^{\dagger}_{b_2j\sigma})$ with $c^{\dagger}_{a_{l}(b_{l})j\sigma}$ creating an electron of spin $\sigma$ at site a$_{l}$ (b$_{l}$) of the $l$th (= 1,2, respectively, for top and bottom) layer of the $j$th unit cell. In momentum-frequency space, $G$ (a 4$\times$4 matrix) can be expressed in terms of the self-energy $\Sigma(k,i\omega_{\ell})$ as
\begin{eqnarray}
G(k,i\omega_{\ell}) = [i\omega_{\ell} +\mu -h_k-\Sigma(k,i\omega_{\ell})]^{-1} \label {Gk}
\end{eqnarray}
with
\begin{eqnarray}
h_k = \begin{pmatrix}
0& \epsilon_k&0&0\\
\epsilon_k^{\ast}&0&-t_1&0\\
0&-t_1&0&\epsilon_k\\
0&0&\epsilon_k^{\ast}&0\\
\end{pmatrix}
\end{eqnarray}
where $\omega_{\ell} = (2\ell+1)\pi T$ is the fermionic Matsubara frequency with $\ell$ as integer number and $T$ the temperature, and $\epsilon_k = -t[1+\exp(-ik_x)+\exp(-ik_y)]$. Here $\mu$ is the chemical potential and is determined by
\begin{eqnarray}
n = \frac{2T}{N_0}\sum_{k\ell}{\rm Tr}G(k,i\omega_{\ell})\exp(i\omega_{\ell}\eta), \label {chm}
\end{eqnarray}
where the factor 2 stems from the spin degeneracy and $\eta$ is an infinitesimally small positive constant. To proceed, we need to provide an approximation for $\Sigma(k,i\omega_{\ell})$. In the following sections, we investigate the possibility of the existence of the GBISP using the SCHFA and the RRDA for the self-energy, respectively.

\section{Studying the existence of the GBISP using the SCHFA}

Let us first consider the physical meaning of the GBISP. As can be seen in Fig. 1, supposing the origin is at the middle point of a b$_1$a$_2$ bond, when changing each atom at site $r_j$ to $-r_j$, the whole lattice is unchanged. This transformation is equivalent to interchanging the top and bottom layers and then rotating the lattice by an angle $\pi$ around the b$_1$a$_2$ bond. In the non symmetry-broken state, the electron system is unchanged with respect to such a transformation. However, when the strong Coulomb interactions drive the system to a GBISP, the two layers cease to be equivalent by inversion symmetry, and the electrons experience different fields on the two layers. Specifically, there may exist net electronic charge accumulation at each atom. We denote the deviations of the electronic charge density from the average value $n$ at each of the four sites of the unit cell as ($\delta_1,\delta_2,-\delta_2,-\delta_1$).
 
\begin{figure} 
\centerline{\epsfig{file=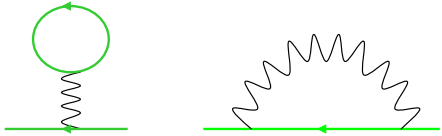,width=7.0 cm}}
\caption{Self-energy of the SCHFA. Left: Hartree term. Right: Fock exchange term. The solid line with an arrow denotes the Green's function. The wavy line is the Coulomb interaction. } 
\end{figure} 

Under the SCHFA or the mean-field approximation, the self-energy is diagrammatically given by Fig. 2. The Hartree term is diagonal, $\Sigma^H_{\mu\nu} = \Delta_{\mu}\delta_{\mu\nu}$, with
\begin{eqnarray}
\Delta_{\mu} &=& \delta_1u_{\mu 1}+\delta_2u_{\mu 2} \label {dmu}\\
u_{\mu 1} &=& \lim_{q\to 0}[v_{\mu 1}(q)-v_{\mu 4}(q)]\label {u1}\\
u_{\mu 2} &=& \lim_{q\to 0}[v_{\mu 2}(q)-v_{\mu 3}(q)]\label {u2}
\end{eqnarray}
where $v_{\mu\nu}(q)$ (with the subscripts $\mu\nu$ being the same as those used in the definition of the Green's function, denoting the four sublattices a$_1$, b$_1$, a$_2$ and b$_2$) is the Fourier component of the Coulomb interaction. In the long-wavelength limit, $v_{\mu\nu}(q)$ behaves like 
\begin{eqnarray}
v_{\mu\nu}(q) \to \frac{2\pi e^2}{S_0\epsilon Q}\exp(-z_{\mu\nu}Q) + \tilde v_{\mu\nu}, ~~~~q \to 0 \label {vq}
\end{eqnarray}
where $S_0 = \sqrt{3}a^2/2$ is the area of the two-dimensional unit cell of monolayer graphene, and $Q$ is the magnitude of the vector $\vec Q = \hat M \vec q$ with\cite{YanB2007} 
\begin{eqnarray}
\hat M = \begin{pmatrix}
1& 0\\
-\frac{1}{\sqrt{3}}&\frac{2}{\sqrt{3}}\\
\end{pmatrix} 
\end{eqnarray}
and where the components of $\vec q$ are along the nonorthogonal axes of the diamond-shaped Brillouin zone. The value of $z_{\mu\nu} = 0$ or $z_0$ (the distance of the two layers) depends on $\mu\nu$ denoting the same layer or two different layers. The last term in Eq. (\ref{vq}), $\tilde v_{\mu\nu}$, is the regular part of the Coulomb potential for $q \rightarrow 0$. The $q$ dependence in Eq. (\ref{vq}) is different from the conventional form because the coordinate axes of the reciprocal lattice where $\vec q$ is defined are nonorthogonal. The wave vector $\vec Q$ is defined in an orthogonal basis.\cite{YanB2007} The relations $\Delta_1 = -\Delta_4$ and $\Delta_2=-\Delta_3$ can be easily checked. 

The Fock exchange term is given by
\begin{eqnarray}
\Sigma^F_{\mu\nu}(k) = -\frac{1}{M}\sum_qv_{\mu\nu}(q)\tilde n_{\mu\nu}(k-q) \label {xc}
\end{eqnarray}
where $M = N_0/4$ is the total number of unit cells in one layer, and $\tilde n_{\mu\nu}(k)$ is given as 
\begin{eqnarray}
\tilde n_{\mu\nu}(k) = T\sum_{\ell}G_{\mu\nu}(k,i\omega_{\ell})\exp(i\omega_{\ell}\eta)-\delta_{\mu\nu}/2, \label {dst}
\end{eqnarray}
which corresponds to the quasiparticle distribution function, the term $-\delta_{\mu\nu}/2$ stemming from the non-normal order of the electronic interaction operator. Under the mean-field approximation, the self-energy $\Sigma_{\mu\nu}(k) = \Sigma^H_{\mu\nu}+\Sigma^F_{\mu\nu}(k)$ is independent of the frequency. By diagonalizing the effective Hamiltonian $h_k + \Sigma (k)$, one can explicitly carry out the frequency summation in Eq. (\ref{dst}). 

The parameters $\delta_1$ and $\delta_2$ are determined by 
\begin{eqnarray}
\delta_1 &=& \frac{1}{M}\sum_k[\tilde n_{11}(k)-\tilde n_{44}(k)], \label {d1}\\
\delta_2 &=& \frac{1}{M}\sum_k[\tilde n_{22}(k)-\tilde n_{33}(k)]. \label {d2} 
\end{eqnarray}
So far, all the components of self-energy and parameters are self-consistently determined by Eqs. (\ref{Gk})-(\ref{d2}). The magnitude of $\delta_1$ is larger than that of $\delta_2$. To see it, consider temporarily the isolated b$_1$ and a$_2$ atoms without Coulomb interaction. Since they are bonded by $t_1$, their atomic degenerate states are split in two bonding-antibonding states with eigenvalue $\pm t_1$. Therefore, the states of the b$_1$ and a$_2$ sublattices contribute mostly to the eigenstates corresponding to the noninteracting energy bands of overall energy separation $\pm t_1$ from the zero energy. At low temperature, the lower band is occupied while the upper band is empty. On the other hand, the valence and conduction bands close to zero energy have eigenvectors which are composed predominantly of the linear combination of atomic states of the a$_1$ and b$_2$ sublattices. The atoms of these two latter sublattices are the first to be affected by the Coulomb interaction, and they are subject to the most charge accumulation in the case of the GBISP. The two parameters $\delta_1$ and $\delta_2$ are not independent, but are correlated through the Green's functions as described by Eqs. (\ref{Gk}) and (\ref{xc})-(\ref{d2}). We can chose $\delta_1$ as the independent order parameter of the GBISP. 
 
\begin{figure} 
\centerline{\epsfig{file=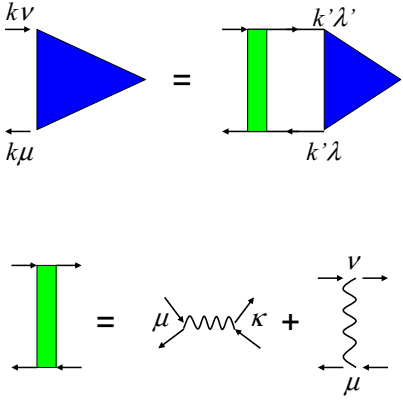,width=6. cm}}
\caption{Equation for the particle-hole propagator $D(k)$. The triangle denotes $D(k)$. The effective interaction between particles and holes is obtained by disconnecting a Green's function line in the self-energy given in Fig. 2. } 
\end{figure} 

To determine the GBISP phase boundary, that is the relation between the critical temperature $T_0$ and the carrier doping concentration $\delta$, we expand the self-energy and the Green's function to first order in the order parameter $\delta_1$. Let us define the matrix 
\begin{eqnarray}
D(k) = \frac{\partial}{\partial\delta_1}[\Sigma(k)-S^{\dagger}\Sigma^{\ast}(k)S]/2 \label{dd}
\end{eqnarray}
with
\begin{eqnarray}
S = \begin{pmatrix}
0& 0&0&1\\
0&0&1&0\\
0&1&0&0\\
1&0&0&0\\
\end{pmatrix}.
\end{eqnarray}
Notice that $\Sigma^{\ast}(k) = \Sigma^{t}(k)$ (the transpose of $\Sigma$) since $\Sigma^{\dagger}(k) = \Sigma(k)$. By this symmetry relation and by the definition in Eq. (\ref{dd}), $D(k)$ has the following structure:
\begin{eqnarray}
D = \begin{pmatrix}
D_{11}& D_{12}&D_{13}&0\\
D^{\ast}_{12}&D_{22}&0&-D_{13}\\
D^{\ast}_{13}&0&-D_{22}&-D_{12}\\
0&-D^{\ast}_{13}&-D^{\ast}_{12}&-D_{11}\\
\end{pmatrix}.
\end{eqnarray}
Therefore, only four elements $D_{11}$, $D_{12}$, $D_{13}$ and $D_{22}$ need to be determined. Under the mean-field approximation, we have the following equation for $D(k)$:
\begin{eqnarray}
D_{\mu\nu}(k) = d_{\mu}\delta_{\mu\nu}-\frac{1}{M}\sum_{k'\lambda\lambda'}v_{\mu\nu}(k-k')f^{\lambda\lambda'}_{\mu\nu}(k')D_{\lambda\lambda'}(k') \nonumber \\ \label {d}
\end{eqnarray}
with  
\begin{eqnarray}
d_{\mu} &=& u_{\mu 1}+u_{\mu 2}\frac{\partial\delta_2}{\partial\delta_1}, \label {u}\\
f^{\lambda\lambda'}_{\mu\nu}(k) &=& T\sum_{\ell}G_{\mu\lambda}(k,i\omega_{\ell})G_{\lambda'\mu}(k,i\omega_{\ell}), \label {f}
\end{eqnarray}
where $G(k,i\omega_{\ell})$'s are the normal-state Green's functions in which $\delta_1 = \delta_2 = 0$. Again, the frequency summation in Eq. (\ref{f}) can be performed analytically. For the normal state, the Green's functions satisfy the relation $G_{\mu\nu} = G_{\bar\nu\bar\mu}$ with $\bar\mu = 5-\mu$. We therefore have $f^{\lambda\lambda'}_{\mu\nu} = f^{\bar\lambda'\bar\lambda}_{\bar\nu\bar\mu}$. By noting these relations, $\partial\delta_2/\partial\delta_1$ can be expressed as
\begin{eqnarray}
\frac{\partial\delta_2}{\partial\delta_1} = \frac{2}{M}\sum_{k\lambda\lambda'}f^{\lambda\lambda'}_{22}(k)D_{\lambda\lambda'}(k).\label{p2}
\end{eqnarray}
Taking the partial derivative of Eq. (\ref{d1}) with respect to $\delta_1$, we obtain the condition for the phase transition, 
\begin{eqnarray}
\frac{2}{M}\sum_{k\lambda\lambda'}f^{\lambda\lambda'}_{11}(k)D_{\lambda\lambda'}(k) = 1. \label{cd}
\end{eqnarray}

Note that $d_{\mu}$ can be formally expressed as 
\begin{eqnarray}
d_{\mu} &=& \frac{2}{M}\sum_{k\lambda\lambda'}[u_{\mu 1}f^{\lambda\lambda'}_{11}(k)+ u_{\mu 2}f^{\lambda\lambda'}_{22}(k)]D_{\lambda\lambda'}(k) \nonumber\\
&=& \frac{2}{M}\sum_{k\kappa\lambda\lambda'}v_{\mu\kappa}(0)f^{\lambda\lambda'}_{\kappa\kappa}(k)D_{\lambda\lambda'}(k), \nonumber
\end{eqnarray}
where in the second line, the definition of $u_{\mu 1(2)}$, $f^{\lambda\lambda'}_{\mu\nu} = f^{\bar\lambda'\bar\lambda}_{\bar\nu\bar\mu}$ and $D_{\bar\lambda\bar\lambda'}(k) = -D_{\lambda'\lambda}(k)$ has been used. (The factor 2 originates from the spin degeneracy.) Putting this result into Eq. (\ref{d}), one obtains the coupled linear equations for $D$'s. The equations are diagrammatically shown in Fig. 3. The function $D(k)$ is actually the particle-hole propagator. The effective interaction between particles and holes is the result of disconnecting a Green's function in the self-energy as given in Fig. 2, by following the procedure explained in Fig. 3. Clearly, these equations are equivalent to solving the problem of a particle-hole propagator with a unity eigenvalue. 

Instead of solving the eigenvalue equations as given in Fig. 3, $D(k)$'s can be determined more easily from Eqs. (\ref{d}), (\ref{u}) and (\ref{p2}) by self-consistent iteration. For a given carrier doping concentration $\delta$, the transition temperature $T_0$ can be found by gradually lowering temperature $T$ from a value higher than the critical one, and solving the equations for $D$'s in the normal state at each step of the process. When the left-hand side of Eq. (\ref{cd}) becomes equal to 1, the critical temperature $T_0$ is reached. 
 
\begin{figure} 
\centerline{\epsfig{file=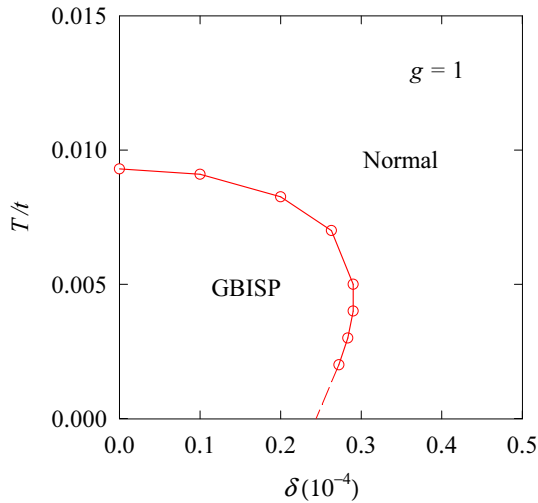,width=7.5 cm}}
\caption{Phase diagram of bilayer graphene in the SCHFA for coupling constant $g = 1$. The symbols are the numerical solution for transition points. The dashed line is an extrapolation of the finite-temperature results to low temperature. } 
\end{figure} 

To numerically solve Eqs. (\ref{Gk}), (\ref{xc}) and (\ref{d})-(\ref{cd}), we need to carefully treat the convolution of the Coulomb interaction $v_{\mu\nu}(q)$ and the function $\tilde n_{\mu\nu}(k-q)$ as appearing in Eq. (\ref{xc}) [and the similar one appearing in Eq. (\ref{d})] because $v_{\mu\nu}(q)$ is singular at $q = 0$. In Appendix A, we present an algorithm to deal with this problem.

In Fig. 4, we show the phase diagram of the electron system in the $\delta-T$ plane for coupling constant $g = 1$. At low temperature and low carrier doping, the system is in the GBISP. The transition temperature as a function of $\delta$ is uniquely defined only at low $\delta < 0.24\times 10^{-4}$. However, in the region $0.24\times 10^{-4} < \delta < 0.3\times 10^{-4}$, each $\delta$ corresponds to two transition temperatures. In the latter case, the phase boundary was determined by adjusting $\delta$ for every given $T$. 

The numerical results for the order parameters $\delta_1$ and $\delta_2$ as functions of $T$ at $\delta = 0$ for coupling constants $g = 0.5$, 1 and 1.8 are shown in Fig. 5. The SCHFA results are denoted as HF. We notice that $|\delta_1| > |\delta_2|$, but $\delta_2$ is not negligibly small, which is different from what has been assumed in the two-band model.\cite{Min} From Fig. 5 one can understand that the charge configuration at the four sites in the unit cell is $(-|\delta_1|,|\delta_2|,-|\delta_2|,|\delta_1|)$ [another solution is $(|\delta_1|,-|\delta_2|,|\delta_2|,-|\delta_1|)$]. The signs of $\delta_1$ and $\delta_2$ are the opposite of each other because with such a charge distribution, the Coulomb interaction between sites a and b in the same plane is attractive and stabilizes the GBISP. It is also seen from Fig. 5 that the transition temperature is higher for a system with stronger coupling. 
 
\begin{figure} 
\centerline{\epsfig{file=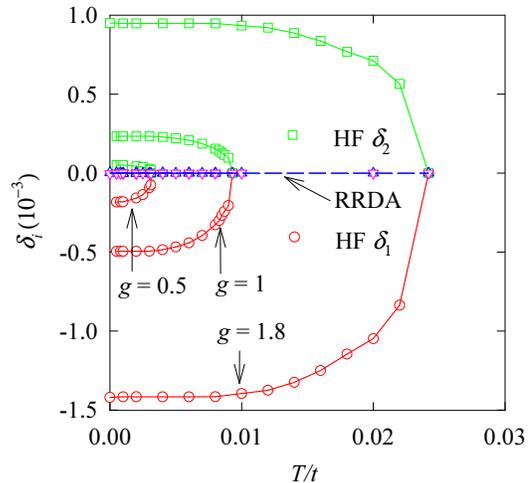,width=7.5 cm}}
\caption{Order parameters $\delta_1$ and $\delta_2$ as functions of temperature $T$ at $\delta = 0$ for coupling constants $g = 0.5$, 1 and 1.8. The symbols refer to numerical results. Circles and squares denoted as HF are the SCHFA results for $\delta_1$ and $\delta_2$, respectively. The RRDA results denoted by triangles ($\delta_1$) and inverse triangles ($\delta_2$) are vanishingly small. } 
\end{figure} 

Our lattice model is different from both the two- and four-band continuum effective models.\cite{Min,MacDonald,McCann2,Novoselov} The two- and four-band continuum models are established under the consideration that the energy scale of quasiparticle spectral resonances involved in the problem is small with respect to a characteristic energy taken from bilayer graphene noninteracting band structure. For the four-band continuum model, the energy should be much less than the bandwidth of the $\pi$ orbitals of graphene. The two-band model for BLG is accurate only in the case in which the quasiparticle energy is much smaller than the gap $t_1$. In the presence of long-range Coulomb interaction $v(q)$, the energy transfer at small $q$ is very large and the assumption for the validity of the two- and four-band continuum models becomes problematic. In this sense, the lattice model appears to be more reasonable. 

Another important difference between the lattice model and the two- and four-band continuum models relates to the valley physics in the Brillouin zone. In the two- and four-band continuum models, the two valleys are independent of each other and the valley index is treated as an overall degeneracy index. On the contrary, within the lattice model two states belonging to different valleys can be connected by nonzero intervalley Hamiltonian matrix elements.

\section{Suppression of the GBISP in the RRDA}

The order parameters $\delta_1$ and $\delta_2$ so obtained by the SCHFA are overestimated because charge fluctuations have been ignored. We here reexamine the possibility of the existence of the GBISP using the RRDA. 

Under the RRDA, besides the Hartree-Fock terms given in Fig. 2, the additional part of the self-energy, denoted by $\Sigma^c(k,i\omega_{\ell})$, is shown in Fig. 6. Each bubble in Fig. 6 is composed of two Green's functions $G$, representing the charge polarizability. In terms of $G$, the elements of $\Sigma^c(k,i\omega_{\ell})$ are expressed as 
\begin{eqnarray}
\Sigma^c_{\mu\nu}(k,i\omega_{\ell}) &=& -\frac{T}{M}\sum_{q,m} G_{\mu\nu}(k-q,i\omega_{\ell}-i\nu_m)W^c_{\mu\nu}(q,i\nu_m)\nonumber
\end{eqnarray}
where $\nu_m$ is the bosonic Matsubara frequency, and $W^c_{\mu\nu}(q,i\nu_m)$ is an effective interaction. The matrix form of $W^c$ is given by
\begin{eqnarray}
W^c(q,i\nu_m) = [1-v(q)\chi(q,i\nu_m)]^{-1}v(q)-v(q) \label {wc}
\end{eqnarray}
with 
\begin{eqnarray}
\chi_{\mu\nu}(q,i\nu_m)=\frac{2T}{M}\sum_{k,\ell} G_{\mu\nu}(k,i\omega_{\ell})G_{\nu\mu}(k-q,i\omega_{\ell}-i\nu_m) \nonumber
\end{eqnarray}
and $v(q)$ is the Fourier component ($4\times 4$ matrix) of the Coulomb interaction. The total self-energy is given by
\begin{eqnarray}
\Sigma_{\mu\nu}(k,i\omega_{\ell})=\Delta_{\mu}\delta{\mu\nu}+ \Sigma^F_{\mu\nu}(k)+ \Sigma^c_{\mu\nu}(k,i\omega_{\ell}).\label{sfe}
\end{eqnarray}
The Green's function $G$ is self-consistently determined and satisfies the microscopic conservation laws.\cite{Baym}

\begin{figure} 
\centerline{\epsfig{file=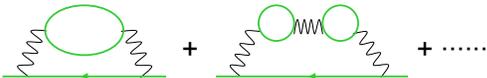,width=7 cm}}
\caption{Additional part of the self-energy besides the Hartree-Fock terms. } 
\end{figure} 

Note that $\Sigma^c$ is a convolution of $G$ and $W^c$, and $\chi$ is a convolution of two $G$'s in momentum and frequency space. The easy way to calculate them is by Fourier transform. At low temperature, the summations index over the Matsubara frequencies should run up to a large frequency cutoff. To reduce the requirement of computer memory storage and accelerating the numerical computation, the special algorithm of Ref. \onlinecite{YanE} can be used.

The interaction $W^c(q,i\nu_m)$ vanishes for $m \to \infty$. For finite $\nu_m$, we need to carefully deal with the singularity at $q = 0$. The Fourier transform $W^c(q,i\nu_m)$ to $W^c(r,i\nu_m)$ is discussed in Appendix B. 

At the ground state for $T = 0$, the Matsubara frequencies $\omega_{\ell}$ and $\nu_m$ are treated as the continuous variables $\omega$ and $\nu$, respectively, and the summations over them are replaced by integrals,
\begin{eqnarray}
T\sum_{\omega_{\ell}}&\to& \int_{-\infty}^{\infty}\frac{d\omega}{2\pi} \nonumber\\
T\sum_{\nu_m}&\to& \int_{-\infty}^{\infty}\frac{d\nu}{2\pi}. \nonumber
\end{eqnarray}

We computed the Green's function within the RRDA. The results for the order parameters $\delta_1$ and $\delta_2$ for $g = 0.5$, 1 and 1.8 at $\delta = 0$ are shown in Fig. 5 and compared with the SCHFA. The doping concentration chosen corresponds to $\delta = 0$, for which the SCHFA transition temperature reaches its maximum. Though $\delta = 0$ is the most favorable case for the GBISP predicted by the SCHFA, the two order parameters are substantially suppressed by CDF in the RRDA; the magnitude of the two parameters is three orders smaller than that of the SCHFA. For inspecting the GBISP ordering at low temperature, the RRDA calculation for $g = 1.8$ is performed down to $T = 0$. From the numerical results, we conclude that there is no the GBISP in systems with $g \le 1.8$ under the RRDA.  

We also computed the Green's function within RPA, in which the polarizability $\chi(q,i\nu_m)$ in $W^c(q,i\nu_m)$ [see Eq. (\ref{wc})] is replaced by the polarizability for noninteracting electrons. At $\delta = 0$, similar to the RRDA, the parameters $\delta_1$ and $\delta_2$ so obtained are vanishingly small. In the RRDA, a replacement of the bare Coulomb interaction in the Hartree term by the screened one is prohibited because where the ring diagrams are equivalent to a self-energy insertion to the Green's function to be renormalized. When performing the RPA calculation, we also need to keep the bare Coulomb interaction in the Hartree term. 

The reason for the suppression of the GBISP under the RRDA is that the exchange interaction is significantly weakened by the screening due to electronic charge-density fluctuations while the Hartree term opposing the charge transfer between the two layers\cite{Min} is not changed. At low temperature, in a wide range of Matsubara frequencies, the exchange interaction is short-ranged and weakened and does not favor the GBISP transition. We point out that the suppression of the GBISP here is not due to prohibition by the Mermin-Wagner theorem.\cite{Mermin} The theorem applies to a system with continuous symmetry; if the symmetry were broken, there would be a logarithmically diverging number of long-wavelength collective fluctuations accompanying the excitations on top of the broken symmetry ground state of the two-dimensional system. In the present case, the inversion is a discrete symmetry operation, and there is no diverging long-wavelength collective fluctuation arising from the breaking of such a symmetry.

\section{Summary}

In summary, we have studied the physics of interacting electrons in bilayer graphene using the lattice model. The possibility of the existence of a GBISP at low temperature and low-carrier-doping concentration is reinvestigated with both the SCHFA and the RRDA. The latter approach takes into account the charge density fluctuations beyond the SCHFA or the mean-field approximation. Under the RRDA, the exchange interaction is weakened substantially, and the existence of a GBISP becomes unsustainable. We have also presented the numerical method for dealing with convolution of a singular Coulomb interaction and the Green's function on the lattice model. This numerical method should be usable for solving problems in many-particle systems.

\acknowledgments
This work was supported by the National Basic Research 973 Program of China under Grants No. 2011CB932700 and No. 2012CB932300, NSFC under Grant No. 10834011, and the Robert A. Welch Foundation under Grant No. E-1146.

\vskip 5mm
\appendix{\bf Appendix A: Calculation of the momentum-space convolution of Coulomb interaction with a smooth function for a lattice model}
\vskip 3mm

For solving problems of two-dimensional electron system in the presence of long-range Coulomb interaction, we sometimes need to deal with the convolution
\begin{eqnarray}
C(k)=\frac{1}{M}\sum_{q} V(q)f(k-q) \label {c}
\end{eqnarray}
where the $q$ summation runs over the first Brillouin zone, $V(q)$ is the Coulomb interaction, and $f(k)$ is a smooth function of $k$. On a lattice, an analytical expression for $V(q)$ is not available but its long-wavelength behavior is known. For the honeycomb lattice, it is given by Eq. (\ref{vq}). $V(q)$ can be divided into long-range and short-range parts. For the honeycomb lattice under consideration, define
\begin{eqnarray}
v^l(q)= \sum_{n} \frac{c}{|\vec Q_n+\vec Q|}\exp(-a_0|\vec Q_n+\vec Q|) \label {vl}
\end{eqnarray}
where $c=2\pi e^2/S_0\epsilon$ is the same factor that appeared in Eq. (\ref{vq}), $\vec Q_n$ is the reciprocal lattice vector, $\vec Q = \hat M \vec q$ is as given in the text, and $a_0$ is an auxiliary parameter. By taking $a_0 = 2a$, the summation in Eq. (\ref{vl}) converges quickly and only a few terms need to be summed up. Clearly, $v^l(q)$ represents a long-range interaction. With $v^l(q)$, $V(q)$ can be written as 
\begin{eqnarray}
V(q) = v^l(q) +v^s(q)
\end{eqnarray}
where $v^s(q)$ is so defined by the equation and is the short-range part of $V(q)$. Note that both $v^l(q)$ and $v^s(q)$ are periodic functions of $q$. Equation (\ref{c}) now is given by
\begin{eqnarray}
C(k)=\frac{1}{M}\sum_{q} v^s(q)f(k-q)+\frac{1}{M}\sum_{q} v^l(q)f(k-q). \label {c1}
\end{eqnarray}
The first integral in Eq. (\ref{c1}) can be safely performed by Fourier transform. In the second integral, the singularity appears at $q = 0$. To find out an auxiliary function for this integral, we pay attention to the expanding form of $f(k-q)$
\begin{eqnarray}
f(k-q) \to f(k) -q_x f_x(k)-q_yf_y(k) \label {fe}
\end{eqnarray}
where $f_{x(y)}(k)= df(k)/dk_{x(y)}$. Define two auxiliary functions $v_x(q)$ and $v_y(q)$ by
\begin{eqnarray}
v_{x(y)}(q)= \sum_{n} \frac{c[q_{x(y)}+(\hat M^{-1} \vec Q_n)_{x(y)}]}{|\vec Q_n+\vec Q|}\exp(-a_0|\vec Q_n+\vec Q|), \nonumber
\end{eqnarray}
where $v_{x(y)}(q)$ is periodic and odd under $\vec q \to -\vec q$. The second integral in Eq. (\ref{c1}) can be written as
\begin{eqnarray}
\frac{1}{M}\sum_{q} v^l(q)f(k-q)& =& \frac{1}{M}\sum_{k'} \{v^l(k-k')[f(k')-f(k)]\nonumber\\
&&+v_x(k-k')f_x(k)\nonumber\\
&&+v_y(k-k')f_y(k)\} \nonumber\\
&&+ f(k)v^l(r)|_{r=0}. \label {c2}
\end{eqnarray}
Now, there is no singularity in the integrand in Eq. (\ref{c2}). The leading term of $v^l(k-k')[f(k')-f(k)]$ as $k' \to k$ is proportional to the derivative of $f$ multiplied with a sign factor since $v^l(k-k') \propto 1/|\hat M(\vec k'-\vec k)|$. This leading term varies discontinuously at $k' = k$. The discontinuity is canceled out by the remaining term $v_x(k-k')f_x(k)
+v_y(k-k')f_y(k)$. As a result, the integrand is a smooth function. The integral can then be carried out numerically. The last term stems from the introduction of the auxiliary functions to the integrand. The value $v^l(r)|_{r=0}$ is given by 
\begin{eqnarray}
v^l(r)|_{r=0} = \frac{1}{M}\sum_{q} v^l(q), \label {vl0}
\end{eqnarray}
which can be calculated explicitly. Replace $q$-summation by 
\begin{eqnarray}
\frac{1}{M}\sum_{q} \to \frac{S_0}{V}\sum_{Q} = S_0\int_{\rm BZ}\frac{d\vec Q}{(2\pi)^2}
\end{eqnarray}
where $S_0 = \sqrt{3}a^2/2$ is the area of the unit cell of the honeycomb lattice as appearing in the text, and BZ means the integral is performed over the first Brillouin zone. The combination of the integration over BZ and the $Q_n$-summation in the definition of $v^l(q)$ equals the integration of the function $c\exp(-a_0Q)/Q$ over the total space of $Q$, 
\begin{eqnarray}\
v^l(r)|_{r=0} &=& S_0\int\frac{d\vec Q}{(2\pi)^2}\frac{c}{Q}\exp(-a_0Q) \nonumber\\
&=& e^2/a_0.
\end{eqnarray}

The function $f(k)$ is assumed to be smooth here. However, for calculating the Fock exchange self-energy, $f(k)$ corresponds to the distribution function and varies dramatically at the Fermi surface at low temperature. In this case, extremely dense grids in a momentum regime covering the Fermi surface should be used to denote the variation of $f(k)$. 

The term $v_x(k-k')f_x(k)+v_y(k-k')f_y(k)$ was introduced in the right-hand side of Eq. (\ref{c2}) in order to smooth the integrand. Because $v_x(q)$ and $v_y(q)$ are periodic and odd functions of $q$, the contribution from the integral of $v_x(k-k')f_x(k)+v_y(k-k')f_y(k)$ is zero. At $T = 0$, there is a discontinuity in $f(k)$ at the Fermi surface and its derivatives $f_x(k)$ and $f_y(k)$ are $\delta$ functions. Therefore, the use of this term at $T = 0$ is unworthy. At $T = 0$, this term should be removed, keeping the discontinuity in the integrand. The cost is to use dense grids near the Fermi surface to ensure the accuracy of the result.

\vskip 5mm
\appendix{\bf Appendix B: Fourier transform of the screening potential $W^c(q,i\nu_m)$}
\vskip 3mm

To take the Fourier transform of $W^c(q,i\nu_m)$ given by Eq. (\ref{wc}) from momentum space to real space, we first pay attention to its singularity at $q = 0$. For small $\nu_m$, because $\chi(q,i\nu_m)$ is finite, the singularity exists only in the second term $v(q)$ on the right-hand side of Eq. (\ref{wc}). Its real space form is known as that given by Eq. (\ref{int}) for its elements. However, at large $\nu_m$, because $\chi(q,i\nu_m)$ is vanishingly small, there is also a singularity in the first term on right-hand side of Eq. (\ref{wc}) and both terms cancel with each other. We need a systematic numerical scheme for the Fourier transform at any $\nu_m$. 

Note that in the limit $q \to 0$, $v(q) \to v_0(q)\hat A$ with $v_0 (q) = c/Q$ (again $c=2\pi e^2/S_0\epsilon$ and $Q = |\hat M \vec q|$) and 
\begin{eqnarray}
\hat A = \begin{pmatrix}
1& 1&1&1\\
1&1&1&1\\
1&1&1&1\\
1&1&1&1\\
\end{pmatrix}. \nonumber
\end{eqnarray}
In the same limit, we have 
\begin{eqnarray}
W^c(q,i\nu_m) &\to& -\frac{\alpha_mc}{Q(Q+\alpha_m)}\hat A  \nonumber\\
&=& W_m(Q)\hat A,\label{wm}
\end{eqnarray}
with 
\begin{eqnarray}
\alpha_m = -c\sum_{\mu\nu}\chi_{\mu\nu}(0,i\nu_m)
\end{eqnarray}
and $W_m(Q)$ so defined by Eq. (\ref{wm}). By observing this asymptotic form, we take the auxiliary function for the Fourier transform as
\begin{eqnarray}
W_a(q) = \sum_nW_m(|\vec Q + \vec Q_n|)\exp(-a_0|\vec Q + \vec Q_n|) \label{wa}
\end{eqnarray}
where $a_0$ again is a parameter for fast convergence of the summation over the reciprocal lattice vectors $\vec Q_n$. The Fourier transform of $W^c(q,i\nu_m)$ is separated into two parts, $[W^c(q,i\nu_m)-W_a(q)\hat A]$ and $W_a(q)\hat A$. There is no singularity in the first one and it can be safely transformed by numerical computation. For the second one, $W_a(q)$ is transformed as
\begin{eqnarray}
W_a(r) &=& a^2\int_{BZ}\frac{d\vec q}{(2\pi)^2}W_a(q)\exp(i\vec q \cdot\vec r) \nonumber\\
&=& S_0\int_{BZ}\frac{d\vec Q}{(2\pi)^2}W_a(q)\exp(i\vec Q \cdot\vec R) \nonumber\\
&=& S_0\int\frac{d\vec Q}{(2\pi)^2}W_m(Q)\exp(i\vec Q \cdot\vec R-a_0Q) \nonumber\\
&=& -\frac{S_0\alpha_mc}{2\pi}\int^{\infty}_0dQ\frac{\exp(-a_0Q)}{Q+\alpha_m}J_0(QR) \label{wi}
\end{eqnarray}
where the first line is the definition with $\vec q$ and $\vec r$ given in the quadrilateral coordinate system, the second line converts $\vec q$ to $\vec Q =\hat M\vec q$ and $\vec R = (\hat M^{t})^{-1}\vec r$ (with $\hat M^{t}$ the transpose of $\hat M$) in the orthogonal systems with $d\vec q = d\vec Q/|\hat M| = \sqrt{3}d\vec Q/2$, the third line comes from the definition of $W_a(q)$ given by Eq. (\ref{wa}), the last line is obtained after the azimuthal integration, and $J_0(QR)$ is the Bessel function. Now the singularity in the integrand exists only when $\alpha_m =0$, but $\alpha_m$ also appears in the front factor and the integral vanishes. However, for large $R$, $J_0(QR)$ oscillates rapidly with $Q$. By observing the large-$QR$ behavior of $J_0(QR)$, we choose the auxiliary function\cite{YanB72} 
\begin{eqnarray}
J_A(z) &=& \sqrt{\frac{1}{\pi z +1}}\{[1+\frac{\pi^2z}{8(\pi z +1)^2}]\sin(z)\nonumber\\
&&+[1-\frac{\pi^2z}{8(\pi z +1)^2}]\cos(z)\}
\end{eqnarray}
and separate $J_0(QR)$ to $J_0(QR)-J_A(QR)$ and $J_A(QR)$. By replacing $J_0(QR)$ with $J_0(QR)-J_A(QR)$ in Eq. (\ref{wi}), the integral can be accurately carried out by simple numerical method. The remaining integral with $J_0(QR)$ replaced by $J_A(QR)$ can be performed using Filon's method.

\end{document}